\documentclass[conference]{IEEEtran}

\usepackage[T1]{fontenc}
\usepackage[utf8]{inputenc}
\usepackage{amsmath,amssymb,amsfonts}
\usepackage{graphicx}
\usepackage{booktabs}
\usepackage{cite}
\usepackage{url}
\usepackage{xcolor}
\graphicspath{{../}}
\definecolor{figred}{rgb}{0.80,0.20,0.20}
\definecolor{figblue}{rgb}{0.10,0.30,0.80}
\definecolor{figgreen}{rgb}{0.05,0.55,0.10}
\definecolor{figorange}{rgb}{0.85,0.45,0.10}
\definecolor{figpurple}{rgb}{0.55,0.10,0.75}

\DeclareRobustCommand{\capsolid}[1]{\mbox{\textcolor{#1}{\rule[0.55ex]{0.95em}{0.9pt}}}}

\title{Self-Calibrated Indoor Tracking from Backscatter Fiducials under NLOS Transmitter Illumination}

\author{
    \IEEEauthorblockN{H\"useyin Yi\u{g}itler, Kalle Ruttik, Jingyi Liao, Alexander Sheverdyaev, and Riku J\"antti}
    \IEEEauthorblockA{
        Department of Information and Communications Engineering\\ Aalto University\\ Espoo, Finland\\ }
    \thanks{This work was supported in part by the EU SNS project Ambient6G under Grant 101192113.}
}

\begin{document}
\maketitle

\begin{abstract}
This paper studies indoor tracking from wall-mounted backscatter fiducials in corridor segments outside direct transmitter illumination. In the measured setup, the transmitter-to-fiducial links are NLOS, whereas the fiducial-to-receiver links along the corridor are largely LOS. The main challenge is that the effective fiducial response is deployment-dependent, so a fixed calibrated link budget is not reliable. We therefore use a grid-based penalized-likelihood tracker that profiles the receiver path, a fitted log-distance slope parameter, and fiducial-specific offsets directly from received powers. The resulting paths can then be reused as surrogate calibration coordinates for residual-map correction, while the same correction with measured calibration coordinates is reported only as a reference. On a short four-fiducial corridor segment, the profiled dual-band tracker gives a 0.52~m median error without measured calibration coordinates, and surrogate residual correction improves this to 0.46~m. With measured calibration coordinates, the same correction and a RADAR-style fingerprint reference both reach 0.31~m. The main remaining limitation is therefore the quality of the surrogate calibration paths rather than the structured observation model itself.
\end{abstract}

\begin{IEEEkeywords}
Indoor positioning, backscatter, dynamic programming, self-calibration, dual-band localization
\end{IEEEkeywords}

\section{Introduction}

Backscatter-based sensing is attractive for indoor positioning because the infrastructure can be reduced to simple, low-power location markers. Wall-mounted backscatter devices can therefore be viewed as backscatter fiducials: fixed landmarks that create position-dependent signatures without active radios at every anchor. This makes it possible to extend tracking into corridor segments outside direct transmitter illumination. The same simplicity also makes localization difficult. Once mounted, a printed monopole no longer has a fixed effective gain: the wall and nearby objects become part of the radiating structure, so the full backscatter link budget varies from one BD location to another~\cite{griffin2009complete}. In practice this appears as anchor-dependent offsets and spatial distortion from mounting, fabrication tolerances, and local multipath. A purely geometric path-loss model is therefore too brittle for accurate tracking from backscatter power alone. Recent real-time ambient-backscatter localization demonstrations show that coarse indoor localization is already feasible from the identities of detected backscatter devices alone~\cite{elsanhoury2025zero}. The present work addresses the complementary problem of extracting finer position information from received power, for which calibration of the fiducial responses becomes important.

The prior work most relevant here lies in three adjacent areas. Classical received signal strength (RSS) fingerprinting methods such as RADAR and Horus rely on site-surveyed radio maps and either direct or probabilistic pattern matching~\cite{bahl2000radar,youssef2005horus}. Later studies improved the maps themselves through interpolation, unsupervised optimization, and data-driven radio-map modeling~\cite{trogh2019unsupervised}. Temporal continuity has been handled with Viterbi or hidden Markov model (HMM) tracking and with particle filters~\cite{trogh2015advanced,jin2018particle}. A separate line on calibration-aware RSS localization and passive-tag or backscatter localization addresses uncertain nuisance parameters and deployment-specific propagation effects~\cite{huang2016rss,romanelli2022robust,zhang2019localizing}. The present work follows the sequential RSS tracking line, with the main emphasis on the observation model and its calibration.

We formulate indoor tracking as penalized-likelihood sequence estimation on a corridor grid around a profiled log-distance tracker. The tracker jointly estimates the trajectory, a fitted log-distance slope parameter, and BD-specific offsets from measured powers alone, and the resulting path can then be reused as a surrogate calibration path for residual-map estimation; the same correction with measured calibration coordinates is reported only as a reference. The framework supports single-band operation at 866~MHz or 899~MHz, dual-band shared-path tracking, and a common-mode bias variant used as an upper-tail regularizer. Relative to earlier RSS trackers, the distinctive elements here are backscatter-specific profiling of unknown BD illumination strength, 
surrogate-path residual estimation built on that profiled stage, and the deployment use of simple backscatter fiducials to cover corridor segments outside direct transmitter illumination.

\section{System Model}

Consider a corridor segment instrumented with $K$ fixed wall-mounted backscatter devices (BDs) at known positions $B_k=[x_k,y_k]^{\top}$, $k=1,\ldots,K$. The receiver follows an unknown discrete-time trajectory $U_t=[u_{x,t},u_{y,t}]^{\top}$, $t=1,\ldots,T$. In the measurement setup, this trajectory is restricted to a corridor section, and the same physical path is observed either at a single carrier or simultaneously at two carriers. In the dual-band case considered here, the bands are 866~MHz and 899~MHz.

For each carrier $f$ and BD $k$, the measured quantity is a received power value in dBm. After optional block averaging in the linear power domain, the observation vector at time $t$ is written as $z_{f,t}=[z_{f,1,t}\ \cdots\ z_{f,K,t}]^{\top}$.
Single-band tracking corresponds to one fixed carrier $f$, while dual-band tracking uses the pair $\mathcal{F}=\{866,899\}$ with a shared hidden receiver path.

The tracking area is discretized into corridor grid cells with centers $\mathcal{U}=\{U_{\ell}\}_{\ell=1}^{L}$.
If the receiver is associated with cell $\ell$ at time $t$, the band-$f$, BD-$k$ observation is modeled as
\begin{equation}
z_{f,k,t}
=
G_{f,k}
+
\Delta_f(U_{\ell},B_k;p_f)
+
r_{f,\ell,k}
+
\varepsilon_{f,k,t},
\label{eq:ipin_obs_model}
\end{equation}
where $p_f$ is a band-dependent log-distance slope parameter, $G_{f,k}$ is a BD-dependent offset, $r_{f,\ell,k}$ is a location-dependent correction term, and $\varepsilon_{f,k,t}$ is the remaining modeling error. To identify $G_{f,k}$ separately from the spatial mean of $r_{f,\ell,k}$, we impose $\sum_{\ell\in\mathcal{V}} a_{\ell} r_{f,\ell,k}=0$ over the valid corridor cells $\mathcal{V}$, with nonnegative weights satisfying $\sum_{\ell\in\mathcal{V}}a_\ell=1$. The deterministic path-loss term is\footnote{This additive form differs from the textbook $-10p\log_{10}(d/d_0)$ convention. Here $d_0$ is used only as a short-range regularizer, while the corresponding reference-power constant is absorbed into $G_{f,k}$. Accordingly, $p_f$ is treated as a fitted slope parameter rather than as a literature path-loss exponent.}
\begin{equation}
\Delta_f(U_{\ell},B_k;p_f)
=
-10p_f\log_{10}\!\left(\|U_{\ell}-B_k\| + d_0\right),
\label{eq:ipin_delta_model}
\end{equation}
with a small offset $d_0>0$ for numerical stability. Because the TX-to-BD illumination path is NLOS in the present setup, $G_{f,k}$ absorbs the unknown per-BD illumination strength created by that forward link, while $p_f$ should be interpreted as a fitted nuisance slope of the mean model rather than as a directly physical propagation exponent.

The key point is that the fixed transmitter location and other slowly varying propagation effects are absorbed into the band- and BD-dependent offsets $G_{f,k}$ and the correction term $r_{f,\ell,k}$. Here $r_{f,\ell,k}$ is a deterministic residual-mean map: for each carrier, grid cell, and BD, it stores the local mean deviation from the profiled log-distance-plus-offset model. It is not a per-sample random perturbation. During cold-start profiling it is held at zero, and during residual-map calibration it is estimated from calibration residuals and then reused for all future samples whose path enters cell $\ell$.

To capture nonuniform model mismatch across the corridor, the residual error in~\eqref{eq:ipin_obs_model} is allowed to have location- and BD-dependent variance, $\varepsilon_{f,k,t}\sim \mathcal{N}(0,\sigma_{f,\ell,k}^2)$, where $\sigma_{f,\ell,k}^2$ is estimated from calibration trajectories. In single-band tracking, each grid cell therefore has band-specific mean entries $\mu_{f,\ell,k}=G_{f,k}+\Delta_f(U_{\ell},B_k;p_f)+r_{f,\ell,k}$ and a corresponding diagonal uncertainty model formed from $\sigma_{f,\ell,k}^2$.

In the dual-band case, the same receiver state $U_t$ explains both carrier measurements, while each carrier keeps its own parameters $\{p_f,G_{f,k},r_{f,\ell,k},\sigma_{f,\ell,k}\}$. The fusion problem is therefore shared-path inference with carrier-specific observation models and a common hidden trajectory.

\section{Method}

\subsection{Dynamic-Programming Corridor Tracker}

The unknown receiver trajectory is represented by the grid-index sequence $\ell_1,\ell_2,\ldots,\ell_T$, where $\ell_t\in\{1,\ldots,L\}$ refers to the corridor grid defined in the system model, and the physical position estimate at time $t$ is $U_{\ell_t}$.

To enforce temporally consistent motion, each cell $\ell$ is connected only to nearby predecessor cells $\mathcal{N}(\ell)=\{\ell' : \|U_{\ell}-U_{\ell'}\|\leq R \}$, where $R$ is the neighborhood radius. The transition penalty between two neighboring cells is $m(\ell',\ell)=\lambda\,\rho_{\delta_m}\!\left(\|U_{\ell}-U_{\ell'}\|\right)$, where $\lambda>0$ controls the strength of motion regularization and $\rho_{\delta_m}(\cdot)$ is the Huber penalty~\cite{huber1992robust}. With additive observation and transition terms, trajectory estimation reduces to a finite-horizon dynamic program on the corridor grid. The resulting objective is therefore a penalized likelihood, equivalently MAP under an unnormalized Huber transition prior.

Assume first that one carrier $f$ is used. Using the calibrated mean and variance terms defined in the system model, the single-band observation cost for sample $t$ and candidate cell $\ell$ is
\begin{equation}
c_t^{(f)}(\ell)
=
\sum_{k=1}^{K}
\left[
\frac{\bigl(z_{f,k,t}-\mu_{f,\ell,k}\bigr)^2}{2\sigma_{f,\ell,k}^2}
+
\log \sigma_{f,\ell,k}
\right].
\label{eq:method_singleband_cost}
\end{equation}
Because the $\log\sigma_{f,\ell,k}$ term is retained, the absolute variance scale matters; the variances are therefore estimated from calibration residuals and floored for numerical stability. Two variance models are used in the experiments. In the simpler constant-variance form, $\sigma_{f,\ell,k}$ is replaced by one pooled value $\sigma_{f,k}$ for each carrier and BD. In the heteroscedastic form, the full per-cell map $\sigma_{f,\ell,k}$ is retained.

When both carriers are available simultaneously, they are processed using a shared hidden path. Each carrier keeps its own calibrated mean and variance maps, but both are required to explain the same grid state $\ell_t$. The dual-band observation cost is
\begin{equation}
c_t^{\mathrm{dual}}(\ell)
=
(1-w)\,c_t^{(866)}(\ell)
+
w\,c_t^{(899)}(\ell),
\label{eq:method_dualband_cost}
\end{equation}
where $0 \leq w \leq 1$ is a relative band weight. In the batch experiments, the profiled dual-band stage first runs a provisional equal-weight pass on the calibration runs and then converts the two bands' robust residual scales into inverse-variance weights; the same rule is reused after surrogate-path or measured-coordinate calibration. This weighted sum is heuristic rather than a full joint Gaussian likelihood. It compensates for unequal band informativeness and residual miscalibration, while the common-mode model below captures only the dominant shared samplewise effect; a full $2K\times 2K$ cross-band covariance remains outside the present model.

To absorb samplewise perturbations that affect both carriers similarly, we also consider a common-mode bias variant. In this case, both carriers share a latent zero-mean Gaussian bias term $b_t$ at sample $t$, so $z_{f,k,t}=\mu_{f,\ell_t,k}+b_t+\varepsilon_{f,k,t}$ with $b_t \sim \mathcal{N}(0,\sigma_b^2)$, where $\sigma_b$ is tuned on the calibration set. This variant is built on the weighted dual-band cost in~\eqref{eq:method_dualband_cost}, not on a full joint Gaussian likelihood. Accordingly, $\alpha_{866}=1-w$ and $\alpha_{899}=w$ are heuristic band-level weights inherited from~\eqref{eq:method_dualband_cost}, whereas $\sigma_{f,\ell,k}^{-2}$ are within-band precision weights. To avoid overloading the residual-map symbol $r_{f,\ell,k}$, define the samplewise innovation for candidate cell $\ell$ as $\eta_{f,k,t}(\ell)=z_{f,k,t}-\mu_{f,\ell,k}$ and let $\omega_{f,\ell,k}=\alpha_f/\sigma_{f,\ell,k}^2$. For fixed $\ell$, the weighted cost is quadratic in $b_t$, so profiling it out gives
\begin{align}
\hat b_{t,\ell}
&=
\frac{\sum_{f\in\mathcal{F}}\sum_{k=1}^{K}\omega_{f,\ell,k}\,\eta_{f,k,t}(\ell)}
{\sum_{f\in\mathcal{F}}\sum_{k=1}^{K}\omega_{f,\ell,k}+\sigma_b^{-2}},
\\
c_t^{\mathrm{dual+com}}(\ell)
&=
\frac{1}{2}\sum_{f\in\mathcal{F}}\sum_{k=1}^{K}\omega_{f,\ell,k}\bigl(\eta_{f,k,t}(\ell)-\hat b_{t,\ell}\bigr)^2
\notag\\
&\quad
+ \sum_{f\in\mathcal{F}}\sum_{k=1}^{K}\alpha_f \log \sigma_{f,\ell,k}
+ \frac{\hat b_{t,\ell}^2}{2\sigma_b^2}.
\end{align}
This is the ``dual-band + common'' variant used in the results section. Thus, $w$ does not replace inverse-variance weighting; it rescales the two carriers before the per-channel precision weights are applied.

Given the observation cost and the transition penalty, the full trajectory is estimated by dynamic programming. The Bellman recursion below propagates the minimum accumulated cost to each state at time $t$. In HMM terminology, the same recursion is the Viterbi algorithm applied to negative log-costs~\cite{forney1973viterbi,trogh2015advanced}. Define the accumulated path cost
\begin{equation}
D_t(\ell)
=
c_t(\ell)
+
\min_{\ell'\in\mathcal{N}(\ell)}
\left[
D_{t-1}(\ell') + m(\ell',\ell)
\right],
\label{eq:method_dp}
\end{equation}
where $c_t(\ell)$ is the single-band cost~\eqref{eq:method_singleband_cost}, the dual-band cost~\eqref{eq:method_dualband_cost}, or the corresponding common-mode cost. Standard backtracking of the minimizing predecessor indices yields the minimum-cost path $\hat{\ell}_1,\hat{\ell}_2,\ldots,\hat{\ell}_T$, and the final position estimate is 
\begin{equation}
    \hat{U}_t = U_{\hat{\ell}_t}.
\end{equation}

\subsection{Observation-Model Calibration}

In the experiments, the observation model is used in three ways. The first is cold-start profiling, where the log-distance slope parameter, BD offsets, and trajectory are estimated from one run without position references. The second is surrogate-path residual correction, where the cold-start tracker is first run on the calibration set and the resulting paths are reused as surrogate calibration positions. The third is the corresponding measured-coordinate residual correction, used only as a reference. For each carrier $f$, let $\mathcal{P}$ denote the candidate set of slope parameters and let $G_f=[G_{f,1}\ \cdots\ G_{f,K}]^{\top}$ collect the per-BD offsets.

In cold-start profiling, a candidate $p\in\mathcal{P}$ defines the mean field $\Delta_f(U_\ell,B_k;p)$ on the grid. The residual-mean map is fixed to $r_{f,\ell,k}=0$ for all $\ell$ and $k$, the BD offsets are initialized from per-BD median powers, and dynamic programming alternates with Huber updates of $G_f$ for up to five outer iterations or until the relative objective change falls below $10^{-3}$. The best candidate is selected by the final robust penalized-likelihood objective. This stage does not require known positions, since the trajectory itself is part of the optimization.

For residual-map correction, a calibration path $U_t$ is provided either by the cold-start tracker or by the measured calibration coordinates. For each $p\in\mathcal{P}$, the corresponding BD offsets are found by robustly fitting the residuals, and the final estimate is obtained as
\begin{equation}
\begin{pmatrix}
\hat{p}_f \\
\hat{G}_f
\end{pmatrix}
=
\underset{\substack{p\in\mathcal{P}\\G_f\in\mathbb{R}}}{\operatorname{arg\,min}}
\sum_{t=1}^{T_{\mathrm{tr}}}\sum_{k=1}^{K}
\rho_{\delta_p}\!\Bigl(
z_{f,k,t}
-\Delta_f(U_t,B_k;p)-G_{f,k}
\Bigr)
\label{eq:method_pg_estimation}
\end{equation}
where $\rho_{\delta_p}(\cdot)$ is the Huber penalty~\cite{huber1992robust}. This correction absorbs the dominant band- and BD-dependent mean effects before any location-dependent correction is estimated.

Once $\hat{p}_f$ and $\hat{G}_{f,k}$ have been obtained, the calibration residuals are $e_{f,k,t}=z_{f,k,t}-\Delta_f(U_t,B_k;\hat{p}_f)-\hat{G}_{f,k}$. Each calibration sample $t$ is then assigned to its nearest grid cell $\ell(t)$. Thus, the time index $t$ always refers to one calibration sample, whereas the cell index $\ell$ refers to the stored map entry that collects all samples with $\ell(t)=\ell$. For notational clarity, define the sample set $\mathcal T_{\ell}=\{t:\ell(t)=\ell\}$ and its cardinality $n_{\ell}=|\mathcal T_{\ell}|$. The raw cellwise residual mean and variance for carrier $f$, cell $\ell$, and BD $k$ are
\begin{align}
\bar{e}_{f,\ell,k}
&=
\frac{1}{n_{\ell}}
\sum_{t\in\mathcal T_{\ell}} e_{f,k,t},
\\
s_{f,\ell,k}^2
&=
\frac{1}{n_{\ell}}
\sum_{t\in\mathcal T_{\ell}} e_{f,k,t}^2
-
\bar{e}_{f,\ell,k}^2.
\end{align}
The variance map entry $\sigma_{f,\ell,k}^2$ is therefore a property of cell $\ell$, obtained from the collection of calibration samples whose estimated or measured path falls into that cell; it is not a time-indexed variance $\sigma_{f,k,t}^2$.

The correction map is then updated from these raw cellwise means. Let $\mathcal{S}_{\lambda_r}(\cdot)$ denote the iterative neighbor smoother used in the implementation: each valid cell is repeatedly replaced by a weighted average of its current value and the mean of its valid neighboring cells, with weights set by the cell sample count and $\lambda_r$. We first smooth the raw mean map,
$\tilde r_{f,\ell,k}
=
\mathcal{S}_{\lambda_r}\!\left(\bar e_{f,\ell,k}\right)$,
and then center it over the valid corridor cells,
\begin{equation}
r_{f,\ell,k}
=
\tilde r_{f,\ell,k}
-
\sum_{j\in\mathcal V} a_j \tilde r_{f,j,k}.
\label{eq:method_r_update}
\end{equation}
Thus, $r_{f,\ell,k}$ is the smoothed local mean residual stored for cell $\ell$ after centering out the constant BD offset, and all later samples visiting $\ell$ reuse that value. The log-variance map is formed analogously from the raw variance statistics, followed by smoothing and clipping for numerical stability. Because the measured path stays close to $x=0$, off-axis map entries are weakly constrained and should be read as local corrections over the corridor strip rather than as fully observed 2-D fields. In Section~V, the same estimator is applied either with the cold-start profiled path or with measured calibration coordinates, so the residual maps become surrogate-path or measured-coordinate corrections depending on the calibration path. The surrogate RADAR reference is built one stage later from those same surrogate paths: from the constant-variance surrogate tracker in single-band and from the surrogate common-mode tracker in dual-band.

The complete processing sequence is summarized below.

\begin{center}
\fbox{\parbox{0.97\columnwidth}{\footnotesize
\begin{tabular}{@{}l@{\hspace{0.5em}}p{0.89\columnwidth}@{}}
1.& Build the corridor grid $\mathcal U$, the neighborhood sets $\mathcal N(\ell)$, and the candidate slope set $\mathcal P$.\\
2.& For each carrier $f$ and candidate $p\in\mathcal P$, initialize $G_f$ from per-BD median powers and alternate dynamic programming for $\ell_{1:T}$ with Huber updates of $G_f$. Keep the candidate with the smallest penalized-likelihood cost.\\
3.& Use the resulting path as either (i) the final cold-start estimate or (ii) a surrogate calibration path. When measured calibration coordinates are available, use them instead.\\
4.& For each carrier, estimate $(\hat p_f,\hat G_f)$ by robust fitting on the chosen calibration path, form cellwise residual mean and variance statistics, smooth them over neighboring cells, and update the residual-mean map $r_{f,\ell,k}$. Use either a pooled constant variance or the full per-cell variance map, depending on the variant.\\
5.& In dual-band operation, estimate the carrier weight from residual scales on the calibration set. If the common-mode variant is used, also tune $\sigma_b$ and evaluate the profiled dual-band common-mode cost.\\
6.& Track the evaluation run by dynamic programming with the calibrated single-band, dual-band, or dual-band common-mode cost.\\
\end{tabular}}}
\end{center}

\subsection{Implementation Parameters}

All table results use the same tracking grid and preprocessing: $x \in [-0.6,0.6]$~m, $y \in [4.75,8.0]$~m, grid spacing $\Delta x=\Delta y=0.4$~m, neighborhood radius $R=0.6$~m, transition weight $\lambda=5$, Huber scale $4$~dB, $p_f\in\{1.2,1.3,\ldots,4.0\}$, five outer profiling iterations, and linear-power block averaging of length $L=5$ before converting back to dBm. For residual-map estimation, the spatial mean and log-variance maps are smoothed with 25 iterations and smoothing weights $\lambda_r=\lambda_{\log\sigma}=4$, with the variance clipped to $[1,12]$~dB. The single-band surrogate-path row uses a constant variance derived from the estimated map, while the measured-coordinate single-band row uses the full heteroscedastic map. In dual-band operation, the common-mode model uses $\sigma_b=4$~dB. The dual-band carrier weight is estimated from inverse robust residual-scale weighting on the calibration set; in the batch comparison this gives $w\approx 0.53$ for the unlabeled and surrogate-path cases and $w\approx 0.56$ for the measured-coordinate case. The RADAR reference uses $k$-nearest-neighbor matching on the block-averaged power vectors, with $k\in\{1,3,5,7,9\}$ selected by leave-one-run-out median-plus-tail score on the calibration split. In the four-BD measured-coordinate residual calibration, $\hat p_{866}=2.6$ and $\hat p_{899}=2.3$, while the calibrated BD offsets spanned approximately $[-107.1,-94.8]$~dB at 866~MHz and $[-103.1,-90.4]$~dB at 899~MHz across the active BDs. The difference between $\hat p_{866}$ and $\hat p_{899}$ mainly indicates band-dependent miscalibration absorbed by the observation model.

\section{Measurement Setup}

\subsection{Environment and Hardware}

The measurements were conducted in an office corridor with mixed LOS/NLOS geometry. The transmitter-to-BD illumination path was NLOS because the transmitters were placed around the corridor corner with no direct path to the wall-mounted BDs, whereas the BD-to-receiver path along the measured segment was largely LOS. This matches the intended deployment setting: tracking in a corridor segment outside direct transmitter illumination using wall-mounted backscatter fiducials. Coordinates are given in meters as $(x,y,z)$. The receiver trolley moved from $(0,0)$ to $(0,14)$ along a straight 14~m path centered between corridor walls at $x=-1$ and $x=1$. We use this route as a controlled proof of concept; despite the simple geometry, the corridor still produces strong multipath and anchor-dependent distortion.

\begin{figure}[t]
\centering
\begin{minipage}[t]{0.55\columnwidth}
\vspace{0pt}
\centering
\includegraphics[width=\linewidth,trim=0.35cm 0.30cm 0cm 0cm,clip]{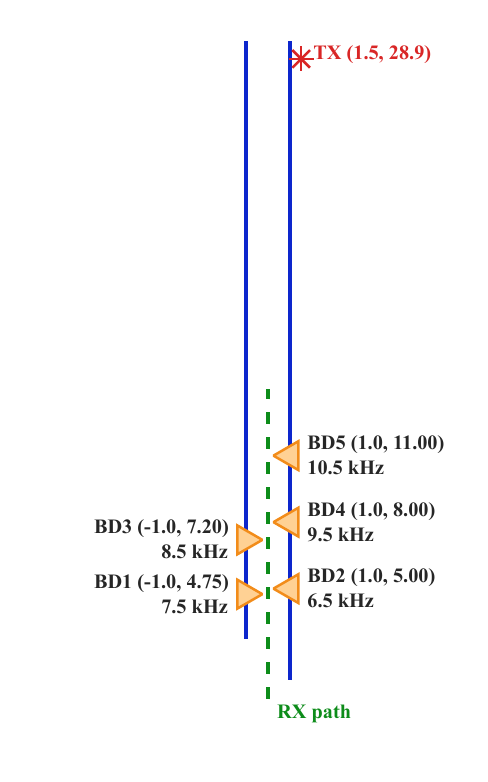}
\end{minipage}\hfill
\begin{minipage}[t]{0.41\columnwidth}
\vspace{0pt}
\centering
\includegraphics[width=0.98\linewidth]{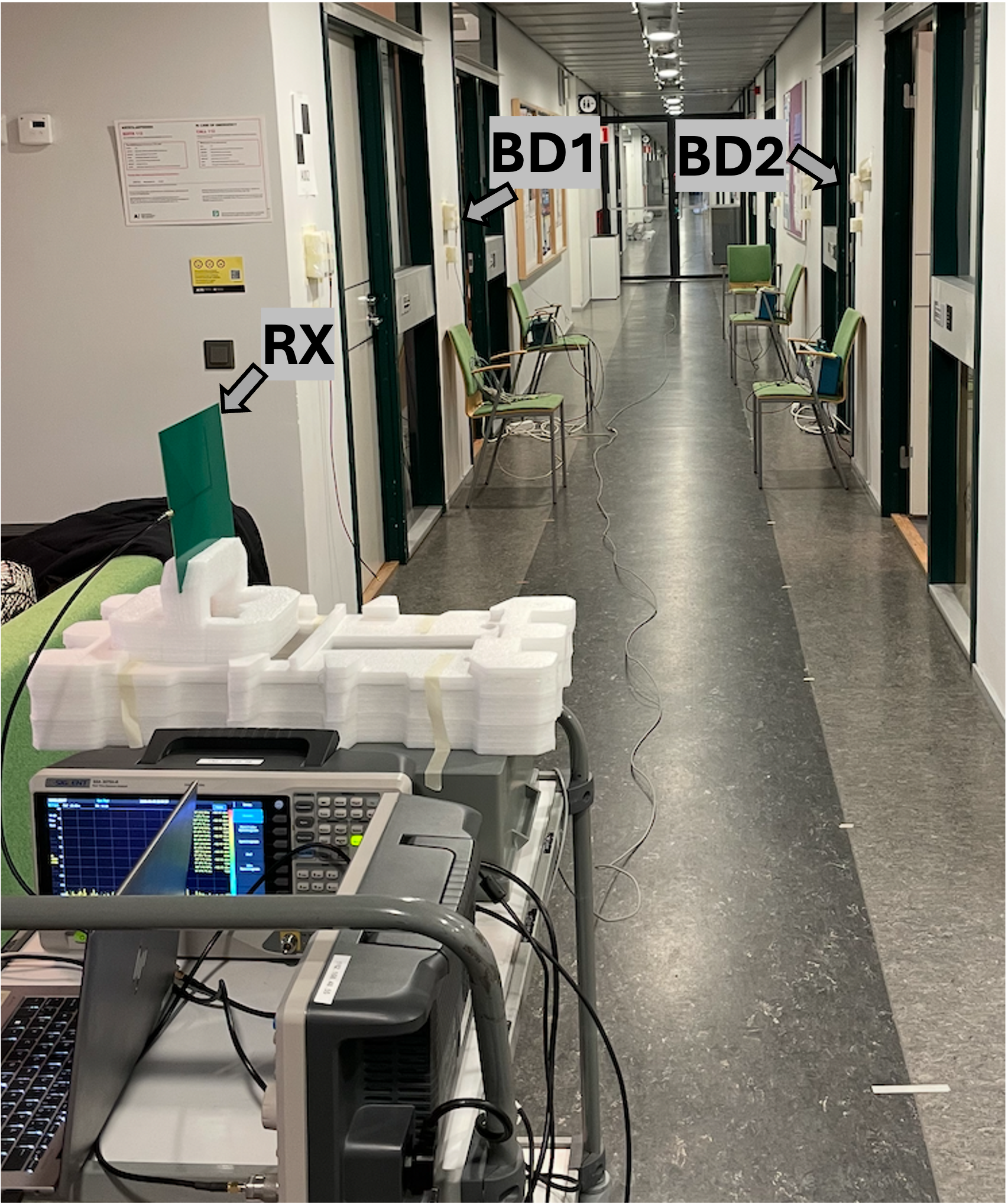}

\vspace{0.2em}

\includegraphics[width=0.98\linewidth]{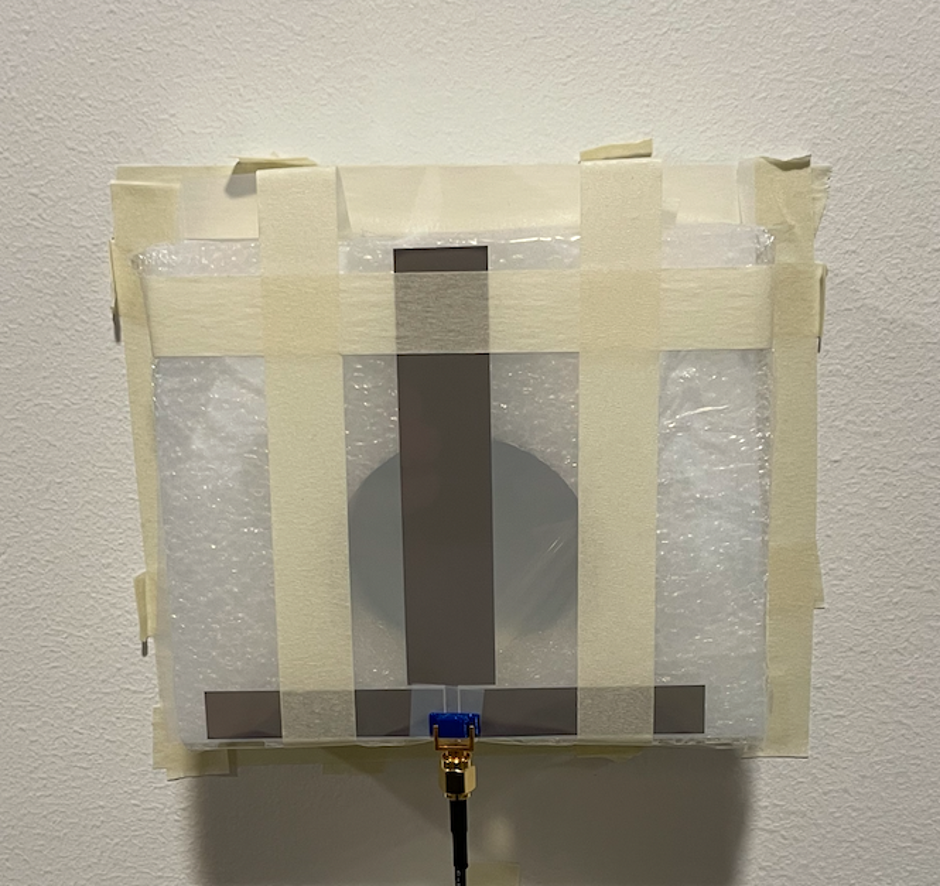}
\end{minipage}
\caption{Measurement setup. Left: top-view geometry. Right: corridor environment and backscatter device antenna.}
\label{fig:meas_setup_photos}
\end{figure}

Two continuous-wave transmit antennas were driven by Rohde \& Schwarz SMBV100A vector signal generators at 30~dBm, and the receiver used a synchronized SIGLENT SSA3075X-R real-time spectrum analyzer with 10~Hz resolution bandwidth. Five BDs were modulated by square waves with 0~V DC offset and 0.75~V amplitude from two Tektronix function generators. The BD/TX positions, BD labels, and switching frequencies are shown in Fig.~\ref{fig:meas_setup_photos}; ST2 drove BD1 and BD3, and ST1 drove BD2, BD4, and BD5.

The batch comparison uses the four-BD subset BD1 to BD4. Raw samples are taken from the corridor interval $y \in [4.0,8.0]$~m, and the tracking grid is restricted to $x \in [-0.6,0.6]$~m and $y \in [4.75,8.0]$~m around the receiver path. Thus, samples in $y \in [4.0,4.75)$~m remain in the table statistics, but their path estimates are constrained by the lower grid boundary at $y=4.75$~m. BD5 lies outside this tracking interval.

\subsection{Measurement Campaigns and Preprocessing}

The data come from two measurement sets, \texttt{s1} and \texttt{s2}, with 32 and 60 continuous runs, respectively. In each run, the same trolley path from $(0,0)$ to $(0,14)$~m was traversed at constant speed.

At 0.609~m/s and a 100~ms sampling interval, successive raw samples correspond to about 6.1~cm along the route. The row index in the measurement matrix is mapped to position by linear interpolation between the start and end points, which implies a practical measured-reference floor of roughly 0.10 to 0.15~m once timing uncertainty and small speed deviations are taken into account. For dual-band processing, the 866~MHz and 899~MHz observations are extracted from the same spectrum-analyzer recording on the same 100~ms sample grid.

The calibration/evaluation protocol is campaign-balanced: the first half of the runs from each campaign are used for calibration and the second half for evaluation, giving 46 calibration runs and 46 evaluation runs. For the table results, adjacent samples are block-averaged over non-overlapping windows of length $L=5$ in the linear power domain before conversion back to dBm; after this preprocessing, the 46 evaluation runs contribute 598 samples over the extracted four-BD interval.

\section{Results}

\subsection{Evaluation Protocol}

Table~\ref{tab:mixed_summary} uses one matched mixed split: 46 calibration runs and 46 evaluation runs over the raw interval $y \in [4.0,8.0]$~m, together with the four-BD tracking grid of Section~IV-A. All methods use the same $L=5$ block averaging and are summarized by Euclidean trajectory error over the resulting 598 evaluation samples. Because the grid starts at $y=4.75$~m, the samples in $y \in [4.0,4.75)$~m are boundary-constrained fringe samples rather than interior-grid samples. The row marked None uses no calibration positions. The surrogate residual rows first estimate surrogate calibration paths by cold-start profiling and then reuse them for residual-map estimation. The surrogate RADAR rows add one more stage and build their reference fingerprint set from the corresponding surrogate residual tracker. Rows marked True use measured calibration coordinates.

\subsection{Matched Batch Comparison}

Table~\ref{tab:mixed_summary} compares the profiled self-calibrated tracker, surrogate-path refinements, and measured-coordinate references under the same split. Single-band cold-start profiling gives 0.51~m median error at 866~MHz and 0.71~m at 899~MHz. With profiled dual-band fusion and an unlabeled variance-based carrier weight estimated from the calibration runs, the median becomes 0.52~m, already improving on the weaker 899~MHz single-band case.

Using the cold-start paths as surrogate calibration positions gives a modest but consistent gain. The medians become 0.50~m at 866~MHz, 0.65~m at 899~MHz, and 0.46~m for the dual-band common-mode tracker. Surrogate RADAR remains competitive, but it is only slightly better at 899~MHz, worse at 866~MHz, and tied on dual-band median while giving a heavier tail, with P90 increasing from 0.85~m to 0.91~m.

With measured calibration coordinates, the remaining gap narrows substantially. The residual-map tracker reaches 0.41~m at 866~MHz, 0.42~m at 899~MHz, and 0.31~m in dual-band, while RADAR also reaches 0.31~m in dual-band and is stronger in the measured-coordinate single-band cases. The structured tracker is therefore most useful in the cold-start and surrogate-path regimes, where it operates without a dense measured radio map. On this route-matched benchmark, the remaining gap is driven mainly by surrogate-path quality.

The dual-band combiner should be read in the same pragmatic spirit. On the profiled calibration runs used for surrogate correction, the cross-band residual correlation is about $0.07$ when pooled over individual BD channels, but about $0.49$ at the samplewise mean-residual level. This points to a mostly samplewise shared disturbance and supports a scalar common-mode stabilizer, while a fuller cross-band covariance model remains future work.

The preprocessing filter has a visible impact. Without block averaging, profiled dual-band degrades from 0.52~m to 0.56~m median and from 0.85~m to 0.98~m at P90, the surrogate dual-band common-mode tracker from 0.46~m to 0.49~m, and the measured-coordinate dual-band common-mode tracker from 0.31~m to 0.39~m. At single-band 899~MHz the median changes little, 0.71~m with $L=5$ versus 0.69~m with $L=1$, but filtering still improves the tail: P95 drops from 1.28~m to 1.13~m. Thus, $L=5$ mainly suppresses short-scale measurement fluctuations and stabilizes the emission cost. The main surrogate dual-band result is also insensitive to a small grid refinement: reducing $\Delta x=\Delta y$ from 0.4~m to 0.3~m changes the median only from 0.462~m to 0.460~m.

\begin{table*}[t]
\centering
\caption{Matched mixed-split trajectory summary over the four-BD evaluation interval extracted from $y \in [4.0,8.0]$~m.}
\label{tab:mixed_summary}
\scriptsize
\setlength{\tabcolsep}{1.6pt}
\begin{tabular}{@{}ll*{15}{r}@{}}
\toprule
Method & Cal. &
\multicolumn{5}{c}{866~MHz} &
\multicolumn{5}{c}{899~MHz} &
\multicolumn{5}{c}{866+899~MHz} \\
\cmidrule(lr){3-7}\cmidrule(lr){8-12}\cmidrule(l){13-17}
& & Med. & P90 & P95 & $|b|$ & $b_y$
& Med. & P90 & P95 & $|b|$ & $b_y$
& Med. & P90 & P95 & $|b|$ & $b_y$ \\
\midrule
Profiled tracking & None
& 0.505 & 0.933 & 1.127 & 0.219 & 0.199
& 0.709 & 1.090 & 1.127 & 0.374 & 0.361
& 0.516 & 0.848 & 1.054 & 0.304 & 0.272 \\
Residual-map tracker & Surr.
& 0.497 & 1.014 & 1.127 & 0.123 & 0.006
& 0.654 & 1.054 & 1.090 & 0.316 & 0.310
& 0.462 & 0.848 & 1.040 & 0.191 & 0.093 \\
RADAR kNN & Surr.
& 0.516 & 1.090 & 1.216 & 0.108 & 0.042
& 0.638 & 1.053 & 1.090 & 0.339 & 0.324
& 0.462 & 0.912 & 1.054 & 0.229 & 0.150 \\
Residual-map tracker & True
& 0.412 & 0.897 & 1.090 & 0.062 & -0.037
& 0.420 & 0.733 & 0.788 & 0.063 & 0.048
& 0.309 & 0.654 & 0.733 & 0.053 & -0.049 \\
RADAR kNN & True
& 0.306 & 0.917 & 1.223 & 0.106 & -0.106
& 0.306 & 0.611 & 1.223 & 0.048 & -0.048
& 0.306 & 0.611 & 0.611 & 0.086 & -0.086 \\
\bottomrule
\end{tabular}
\vspace{5pt}

\footnotesize\noindent All rows use 46 calibration runs, 46 evaluation runs, and $L=5$ block averaging. Surr.\ denotes surrogate-path calibration, True measured-coordinate calibration, and the last two columns report pooled $|b|$ and $b_y$.
\end{table*}

\begin{figure}[t]
\centering
\includegraphics[width=\linewidth,trim=0 0 0 1cm,clip]{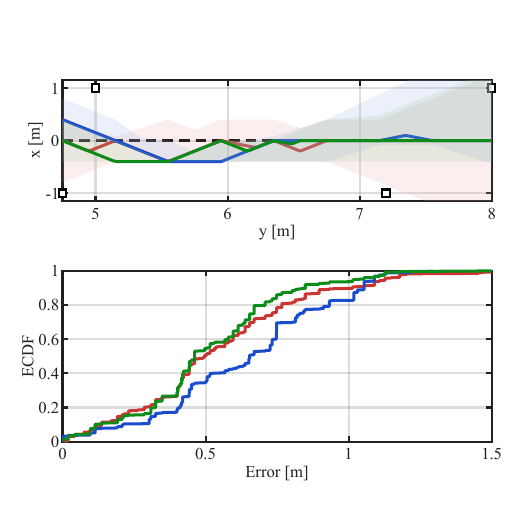}
\caption{Compact mixed-split summary. Top: median surrogate-path trajectories $\capsolid{figred}$ 866~MHz, $\capsolid{figblue}$ 899~MHz, and $\capsolid{figgreen}$ 866+899~MHz, plotted as lateral offset $x$ versus along-track coordinate $y$, with same-color filled 5th to 95th percentile bands; dashed black is the reference path at $x=0$, and open squares mark the BD positions. Bottom: ECDFs of the same three surrogate modes, again $\capsolid{figred}$ 866~MHz, $\capsolid{figblue}$ 899~MHz, and $\capsolid{figgreen}$ 866+899~MHz.}
\label{fig:pseudo_summary}
\end{figure}

Figure~\ref{fig:pseudo_summary} summarizes the surrogate comparison. The upper panel plots lateral offset $x$ against along-track coordinate $y$ and shows the median surrogate paths at 866~MHz, 899~MHz, and in dual-band, with filled 5th to 95th percentile bands. The lower panel shows the corresponding ECDFs; RADAR is left to Table~\ref{tab:mixed_summary} to keep the figure readable. Dual-band surrogate tracking is the strongest proposed mode: its median path stays closest to the corridor centerline, whereas the 899~MHz path bends toward positive $x$ near the upper end, consistent with the larger bias below. The percentile bands are tight in the middle and widen near the ends, where the geometry is weaker.

\subsection{Bias and Dual-Band Effect}

The bias columns in Table~\ref{tab:mixed_summary} show the same pattern as the ECDFs. Cold-start profiling has a clear positive along-track drift, from $|b|=0.22$~m and $b_y=0.20$~m at 866~MHz to 0.37~m and 0.36~m at 899~MHz. Dual-band profiling lowers the median error but still keeps $|b|=0.30$~m and $b_y=0.27$~m. Surrogate correction helps most at 866~MHz and in dual-band, reaching $|b|=0.12$~m with $b_y\approx 0$ at 866~MHz and 0.19~m with 0.09~m in dual-band, whereas the 899~MHz surrogate rows remain clearly biased. With measured calibration coordinates, bias becomes small for both the residual-map tracker and RADAR; in dual-band the residual tracker is slightly less biased than RADAR, even when their medians are similar.

Overall, dual-band processing remains the strongest of the proposed modes. Cold profiling improves from 0.71~m at 899~MHz to 0.52~m in dual-band, and surrogate common-mode residual tracking reaches 0.46~m with a lighter tail than surrogate RADAR. With measured calibration coordinates, the dual-band residual tracker reaches 0.31~m and is essentially tied with RADAR on median error. The remaining gap again points mainly to surrogate-path quality. These results come from a short straight corridor segment and should be read as a controlled proof of concept rather than as evidence of general 2-D tracking coverage. Even so, all dual-band reference rows remain well below the indoor Ambient IoT tracking target in 3GPP TS~22.369, which specifies 1~m to 3~m positioning accuracy with 90\% positioning-service availability for indoor tracking~\cite{3gpp22369}.

\section{Conclusion}
This paper presented a self-calibrated profiled log-distance tracker for indoor corridor tracking using backscatter-fiducial power measurements. The results show that wall-mounted backscatter fiducials can extend positioning coverage beyond direct transmitter illumination when their deployment-dependent offsets are estimated jointly with the path. On the studied four-BD interval, the proposed tracker achieves sub-metre median accuracy without measured calibration coordinates, and surrogate-path residual correction narrows the gap to a calibrated RADAR benchmark.

The findings indicate that the main remaining limitation is surrogate-path quality rather than the structured observation model itself. In particular, iterating between path estimation and residual-map calibration is a natural way to reduce surrogate-path bias. They also suggest that the proposed method is promising for indoor positioning in obstructed or non-line-of-sight transmitter–receiver conditions. Open questions include path diversity, parameter sensitivity, residual-map transferability, and explicit cross-band covariance modeling.

\vskip 32pt
\bibliographystyle{IEEEtran}
\bibliography{./references}

@inproceedings{elsanhoury2025zero,
  title={Zero-energy devices for {6G}: First real-time indoor localization thanks to ambient backscattering of commercial {4G} {UEs}},
  author={ElSanhoury, Ahmed and Galal, Islam and AlKady, Khaled and ElKhodary, Aml and Elbiali, Hashem and Phan-Huy, Dinh-Thuy and Hassan, Ayman M},
  booktitle={2025 Joint European Conference on Networks and Communications \& {6G} Summit ({EuCNC/6G Summit})},
  pages={235--240},
  year={2025},
  organization={IEEE}
}

@incollection{huber1992robust,
  title={Robust estimation of a location parameter},
  author={Huber, Peter J.},
  booktitle={Breakthroughs in Statistics: Methodology and Distribution},
  pages={492--518},
  year={1992},
  publisher={Springer}
}

@inproceedings{bahl2000radar,
  title={RADAR: An in-building {RF}-based user location and tracking system},
  author={Bahl, Paramvir and Padmanabhan, Venkata N.},
  booktitle={Proceedings IEEE INFOCOM 2000},
  volume={2},
  pages={775--784},
  year={2000},
  organization={IEEE}
}

@inproceedings{youssef2005horus,
  title={The {Horus} {WLAN} location determination system},
  author={Youssef, Moustafa and Agrawala, Ashok},
  booktitle={Proceedings of the 3rd International Conference on Mobile Systems, Applications, and Services},
  pages={205--218},
  year={2005}
}

@article{griffin2009complete,
  title={Complete link budgets for backscatter-radio and {RFID} systems},
  author={Griffin, Joshua D and Durgin, Gregory D},
  journal={IEEE Antennas and Propagation Magazine},
  volume={51},
  number={2},
  pages={11--25},
  year={2009},
  publisher={IEEE}
}

@inproceedings{jin2018particle,
  title={Towards Scalable Indoor Localization with Particle Filter and {Wi-Fi} Fingerprint},
  author={Jin, Feiyu and Liu, Kai and Zhang, Hao and Feng, Liang and Chen, Chao and Wu, Weiwei},
  booktitle={2018 15th Annual IEEE International Conference on Sensing, Communication, and Networking (SECON)},
  pages={1--2},
  year={2018},
  organization={IEEE}
}

@article{trogh2015advanced,
  title={Advanced real-time indoor tracking based on the {Viterbi} algorithm and semantic data},
  author={Trogh, Jens and Plets, David and Martens, Luc and Joseph, Wout},
  journal={International Journal of Distributed Sensor Networks},
  volume={11},
  number={10},
  pages={271818},
  year={2015},
  publisher={SAGE Publications}
}

@article{trogh2019unsupervised,
  title={An unsupervised learning technique to optimize radio maps for indoor localization},
  author={Trogh, Jens and Joseph, Wout and Martens, Luc and Plets, David},
  journal={Sensors},
  volume={19},
  number={4},
  pages={752},
  year={2019},
  publisher={MDPI}
}

@article{forney1973viterbi,
  author={Forney, G. David},
  title={The {Viterbi} Algorithm},
  journal={Proceedings of the IEEE},
  year={1973},
  volume={61},
  number={3},
  pages={268--278},
  doi={10.1109/PROC.1973.9030}
}

@article{huang2016rss,
  title={{RSS}-based method for sensor localization with unknown transmit power and uncertainty in path loss exponent},
  author={Huang, Jiyan and Liu, Peng and Lin, Wei and Gui, Guan},
  journal={Sensors},
  volume={16},
  number={9},
  pages={1452},
  year={2016},
  publisher={MDPI}
}

@misc{3gpp22369,
  author = {{3GPP}},
  title = {{TS} 22.369: Service requirements for Ambient power-enabled {IoT}},
  year = {2024},
  howpublished = {\url{https://portal.3gpp.org/desktopmodules/Specifications/SpecificationDetails.aspx?specificationId=4232}},
  note = {Release 19, under change control; accessed 2026-04-22}
}

@article{romanelli2022robust,
  title={Robust simultaneous localization and mapping using range and bearing estimation of radio ultra high frequency identification tags},
  author={Romanelli, Fabrizio and Martinelli, Francesco and Di Giampaolo, Emidio},
  journal={IEEE Transactions on Control Systems Technology},
  volume={31},
  number={2},
  pages={772--785},
  year={2022},
  publisher={IEEE}
}

@inproceedings{zhang2019localizing,
  title={Localizing backscatters by a single robot with zero start-up cost},
  author={Zhang, Shengkai and Wang, Wei and Tang, Sheyang and Jin, Shi and Jiang, Tao},
  booktitle={2019 IEEE Global Communications Conference (GLOBECOM)},
  pages={1--6},
  year={2019},
  organization={IEEE}
}

\end{document}